\documentclass[runningheads]{llncs}
\usepackage{graphicx}
\usepackage{subfiles}
\usepackage{amsmath}
\usepackage{latexsym}
\usepackage{listings}
\usepackage{courier}
\usepackage{tcolorbox}
\usepackage{cite}

\usepackage{float}

\usepackage{fancyvrb}
\usepackage{fvextra}

\tcbset{colback=white!0!white}  
\tcbsetforeverylayer{colframe=blue!0!black} 

\def\uppaal{\raisebox{.0ex}{\scshape Uppaal}}  
\def\tockCSP{{\slshape tock-CSP}}

\begin{document}

\title{Temporal Reasoning Through Automatic Translation of tock-CSP into Timed Automata  \thanks{The authors gratefully acknowledge the financial support of Petroleum Technology Development Fund (PTDF)}
}


\author{Abdulrazaq Abba \inst{1,2} \orcidID{0000-0003-4465-3858} \and
Ana Cavalcanti \inst{1}
\and
Jeremy Jacob \inst{1}
}

\authorrunning{A. Abba, et al}



\institute{University of York, UK \\ \email{aha526@york.ac.uk}\and
Bayero University, Kano, Nigeria   \\ \email{ahabba.csc@buk.edu.ng}
}

\maketitle              

\begin{abstract}
In this work, we consider translating \tockCSP{} into Timed Automata for \uppaal{} to facilitate using \uppaal{} in reasoning about temporal specifications of \tockCSP{} models. The process algebra \tockCSP{} provides textual notations for modelling discrete-time behaviours, with the support of tools for automatic verification. Similarly,  automatic verification of Timed Automata (TA) with a graphical notation is supported by the \uppaal{} real-time verification toolbox \uppaal. The two modelling approaches, TA and \tockCSP{}, differ in both modelling and verification approaches, temporal logic and refinement, respectively, as well as their provided facilities for automatic verification. For instance, liveness requirements are difficult to specify with the constructs of \tockCSP{}, but they are easy to specify and verify in \uppaal{}. To take advantage of temporal logic, we translate \tockCSP{} into TA for \uppaal;  we have developed a translation technique and its supporting tool. We provide rules for translating \tockCSP{} into a network of small TAs for capturing the compositional structure of \tockCSP{} that is not available in TA. For validation, we start with an experimental approach based on finite approximations to trace sets. Then, we explore mathematical proof to establish the correctness of the rules for covering infinite traces.

\keywords{Translation  \and \tockCSP{} \and Timed-Automata}
\end{abstract}

\section{Introduction}   

Communicating Sequential Processes (CSP) is an established process algebra that provides a formal notation for both modelling and verifying concurrent systems \cite{R10, S10, H78}. The use of CSP for verification has been supported by several tools including powerful model-checkers \cite{TP14, R10, SL09}. 


Interest in using CSP motivated \cite{R10} the introduction of  support for modelling \emph{discrete} timed systems; \tockCSP{} provides an additional event $tock$ to record the progress of time. The notation of \tockCSP{} is encoded in that of CSP, thus, verification is supported by the existing tools. As a result, \tockCSP{} has been used to verify real-time systems, such as security protocols \cite{NS00} and railway systems \cite{YF12}. Recently, \tockCSP{} has been used to capture the semantics of RoboChart, a domain-specific language for modelling robotics applications \cite{MR16}.



In this work, we present a technique for automatic translation of \tockCSP{} into TA to enable using \uppaal \cite{BD06} and temporal logic to verify \tockCSP{} models. \uppaal{} is a tool-suite for modelling hybrid systems using a network of TAs and verifying the systems. We describe list of translation rules and their implementation into a tool.


Both temporal logic and refinement are powerful approaches for model checking \cite{G08}.  The refinement approach models both the system and its specifications with the same notation \cite{R10, S10}, while temporal logic enables asking whether a system captures logical formul\ae{} of the requirements specification in the form of $system \models formula$ \cite{CE86}. 


Lowe has investigated the relationship between the refinement approach (in CSP) and the temporal logic approach \cite{G08}. The result shows that, in expressing temporal logic checks using refinement, it is necessary to use the infinite refusal testing model of CSP. The work highlights that capturing the expressive power of temporal logic to specify the availability of an event (liveness specification) is not possible in the refinement model. Due to the difficulty of capturing refusal testing, automatic support becomes problematic. A previous version of FDR supports refusal testing, but not its recent more efficient version \cite{TP14}.


Additionally, Lowe's work \cite{G08} proves that simple refinement checks cannot match the expressive power of temporal logic, especially the three operators: $eventually$  ($\diamond p$: $p$ \textit{will hold in a subsequent state}),  $until$ ($p  \mathcal{U} q$: \textit{$p$ holds in every state until $q$ holds})  and their $negations$:  ($\neg ( \diamond p$)  and $\neg (p  \mathcal{U} q$). These three operators express behaviour captured by infinite traces. Our work presented here facilitates checking such specifications.



\begin{example} 
\label{ads}
Consider an Automatic Door System (ADS) that opens a door, and after at least one-time unit, closes the door in synchronisation with a lighting controller, which turns off the light. In \tockCSP{}, this is expressed as:

\begin{lstlisting}
       ADS = Controller [|{close}|] Lighting
Controller = open -> tock -> close -> Controller
  Lighting = close -> offLight -> Lighting
\end{lstlisting}

\end{example}



\hspace{-0.7cm}
ADS has two components \textemdash{} \lstinline{Controller} and \lstinline{Lighting} \textemdash{} that synchronise on the event \lstinline{close}, which enables \lstinline{Lighting} to turn off the light after closing the door. In  \tockCSP{}, there is no direct way of checking if the system eventually turns off the light. However, temporal logic provides a direct construct for specifying liveness requirements, supported in \uppaal{}, as follows. 

    
\vspace{-0.22cm}
\begin{itemize}
    \item \lstinline{A<> offLight}   \hspace{0.3cm}  - - \textit{The system eventually turns off the light
        }
        
\end{itemize}

\noindent
\uppaal{} uses a subset of Timed Computational Tree Logic (TCTL) based on the notions of path and state \cite{BD06}.  A path formula quantifies over paths (traces), whereas a state formula describes locations. There are different forms of path formul\ae{}. Liveness is either \lstinline{A<>q} \textit{(q is eventually satisfied)} or \lstinline{p --> q} \textit{(a state satisfying p leads to a state satisfying q)}. A reachability formula in form of \lstinline{E<>q} \textit{(a state satisfying $q$ is reachable from the initial state)}. Safety is expressed as either \lstinline{A[]q}  \textit{(q holds in all reachable states)} or \lstinline{E[]q} \textit{(q holds in all states on at least one path)}.

To verify the correctness of the translation technique, first, we construct a systematic list of interesting \tockCSP{}  processes, which pair all the constructs of \tockCSP{} within the scope of this work. Second, we use the developed translation technique and its tool to translate the formulated processes into TA for \uppaal.  Third, we use another tool we have developed to generate and compare finite traces of the input \tockCSP{} models and the traces of the translated TA models.


We use Haskell \cite{H16}, a functional programming language, to express, implement and evaluate the translation technique. The expressive power of Haskell helps us provide formal descriptions of the translation technique as a list of translation rules, which is also suitable for constructing mathematical proof.


The structure of this paper is as follows. Section \ref{background} provides the essential background material.  In Section \ref{trans_rule}, we summarise the translation technique. We discuss an evaluation of the translation technique in Section \ref{evaluation}. In Section \ref{related_work}, we highlight related works and present a brief comparison with this work. Finally, in Section \ref{conclusion}, we highlight future extensions of this work and conclude. 
Additional details of the missing proofs, implementation and further examples can be found in the extended versions \cite{thesis, extendedVersion}.



\section{Background}
\label{background}



As an extension of CSP, \tockCSP{} provides notations for modelling processes and their interactions, such as the basic processes: \lstinline{SKIP} and \lstinline{STOP}, for successful termination and deadlock, respectively. Operators include \lstinline{prefix (->)} for describing availability of an event. For example, the process \lstinline{move->SKIP} represents a mechanism that moves once and then terminates.


There are binary operators such as sequential composition (;), which combines two processes serially. For instance, the process \texttt{P3 = P1;P2} behaves as process \texttt{P1}, and after successful termination of  \texttt{P1}, then \texttt{P3} behaves as \texttt{P2}. There are other binary operators for concurrency, choice and interruption. Also, CSP has a special event $tau \ ( \tau ) $ for invisible actions that are internal to a system. The collection of these operators provides a rich set of constructs for modelling untimed systems  \cite{R10, S10}. 

For modelling time, \tockCSP{} has an event \texttt{tock} \cite{R10}, which specifies at least single unit of time. For example, the following process \texttt{Pt} specifies behaviour that \lstinline{moves} and then after at least two time units, \lstinline{turns} and terminates.
\vspace{-0.21cm}
\begin{equation*}
\texttt{Pt = move->tock->tock->turn->SKIP}
\end{equation*}

Timed Automata for \uppaal{} model hybrid systems as a network of TA. Mathematically, a TA is a tuple $(L, l_0, C, A, E, I)$, where $L$ is a set of locations such that $l_0$  is the initial location, $C$ is a set of clocks, $A$ is a set of actions, $E$ is a set of edges that connects the locations $L$, and $I$ is an invariant associated to a location $l \in L$ in the form of $I:L \longrightarrow B(C)$. So, edges $E \subseteq (L \times A \times B(C) \times 2^C \times L)$ from a location $l \in L$ triggered by an action $a \in A$, guarded with a guard  $ g \in B(C)$, and associated clock $ c \in C$  that is reset on following the edge to a location $l \in L$ \cite{BD06, B11}. 


A system is modelled as a network of TAs that communicate via either synchronous channel communication or shared variables. A sending channel is decorated with an exclamation mark $(c!)$ while the corresponding receiving channel is decorated with a question mark $c?$. A TA performs an action $c!$ to communicate with another TA that performs the corresponding co-action $c?$. There are also broadcast channels for communication among multiple TAs, in the form of one-to-many communications (one sender with multiple receivers).


For expressing urgency, there are urgent channels and urgent locations that do not allow delay. There are also committed locations; urgent locations that must participate in the next transition, which is useful for expressing atomicity; a compound actions spanning multiple transitions that must be executed as a unit. Invariants specify precise delay and enforce progress \cite{BD06}. In  \uppaal{}, networks of TAs model  system's components and an explicit operating environment \cite{B09, LP07, BD06}.





\section{An Overview of the Translation Technique}

\label{trans_rule}

Our translation technique takes an input \tockCSP{} model and produces a list of TAs. The occurrence of each \tockCSP{} event is captured in a small TA with an \uppaal{} action, which records an occurrence of the translated event. The small TAs are composed into a network of TAs that capture the behaviour of the input \tockCSP{} model. The network of small TAs give us enough flexibility to capture the compositional structure of \tockCSP{}.

 \begin{figure} 

\centering

\includegraphics [width=0.7\textwidth, height=0.4\textwidth]{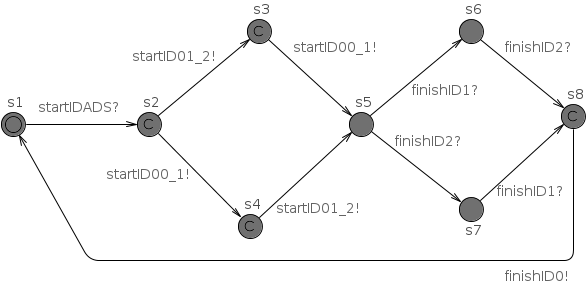}  \\

(a) TA00

\begin{tabular}{c c}

\includegraphics [width=0.4\textwidth]{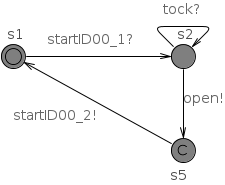} &

\includegraphics [width=0.4\textwidth]{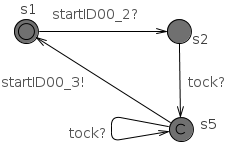} \\

 (b) TA01 &  (c) TA02 \\

\includegraphics [width=0.55\textwidth]{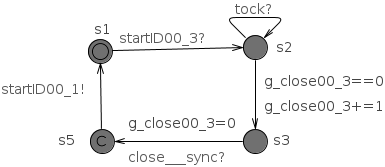} &

\includegraphics [width=0.4\textwidth]{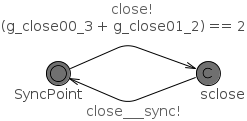} \\

(d) TA03 & (e) TA04 \\

\includegraphics [width=0.5\textwidth]{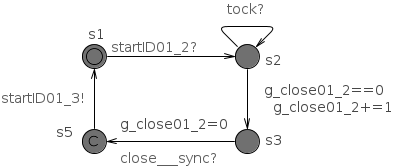} &

\includegraphics  [width=0.4\textwidth]{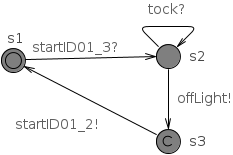}  \\

(f) TA05 &  (g) TA06 \\



\end{tabular}

\caption{A list of networked TA for the translation of the process ADS.}
\label{ta_ads}
\end{figure}

\begin{figure}

    \centering

\includegraphics [width=0.8\textwidth]{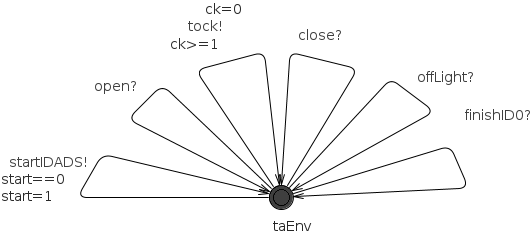}

\caption{An environment TA for the translated behaviour of the process ADS.}

\label{ads_env}
\end{figure}

\begin{example} A translation of the process ADS, from Example \ref{ads}, produces a network of small TAs in Figure \ref{ta_ads}. TA0 captures concurrency by starting the two automata for the processes \lstinline{Controller} and \lstinline{Lighting} in two possible orders \textemdash{} either \lstinline{Controller} then \lstinline{Lighting} or vice versa \textemdash{} depending on the operating environment. Here we use the committed locations (\lstinline{s2}, \lstinline{s3} and \lstinline{s4}) to show that starting the concurrent automata is a compound action. Then TA0 waits on state \lstinline{s5} for the termination actions in the two possible orders. For the termination, we do not use committed locations because the processes can terminate at a different times. TA0 synchronises the processes before terminating the system with the action \lstinline{finishID0}.

TA01, TA02 and TA03 capture the behaviour of the process \lstinline{Controller}. TA01 captures the occurrence of the event \lstinline{open}. TA02 captures the occurrence of  \lstinline{tock?} to synchronise with the environment TA in recording the progress of time.  TA03 captures the event \lstinline{close} to synchronise with the controller TA04.

TA05 and TA06 capture the behaviour of the process \lstinline{Lighting}. TA05 captures \lstinline{close}, which also synchronises with TA04. Then, TA06 captures the event \lstinline{offLight}. Finally, Figure \ref{ads_env} is the environment TA that has co-actions for all the translated events. Also, the environment TA serves the purpose of `closing’ the overall system as required for the model checker. In the environment TA, we use the variable \lstinline{start} to construct a guard \lstinline{start==0} that blocks the environment from restarting the system. 

\end{example}




The main reason for using a list of small TAs is to capture the compositional structure of \tockCSP{}, which is not available in TA \cite{DH08}. For instance, it can be argued that a linear process constructed with a series of prefix operators can be translated into a linear TA. However, the compositional structure of \tockCSP{} is not suitable for this straightforward translation. For instance, consider a case where the linear process is composed of an interrupting process, the behaviour is no longer linear because the process can be interrupted at any stable state, as illustrated in Example \ref{egPI}. Also, this problem can be seen in translating a process like \lstinline{P = (e1->SKIP)[]((e2->SKIP)|||(e3-SKIP))}, which contains both external choice and concurrency. However, a network of small TAs provides enough flexibility for composing TA in various ways to capture the behaviour of the original \tockCSP{} process. 

In constructing the networked TAs, we use additional \textbf{\textit{coordinating actions}} to link the list of small TAs to establish the flow of the input \tockCSP{} model. For example, the channel \lstinline{startIDADS} links the environment TA (Figure \ref{ads_env}) with TA00 (Figure \ref{ta_ads}), on performing the action \lstinline{startIDADS!} and its co-action \lstinline{startIDADS?}.  A precise definition of the coordinating action is as follows.





\begin{definition}

 \textbf{A Coordinating Action} is an \uppaal{} action that does not correspond to a \tockCSP{} event. There are six types of coordinating actions: \textbf{Flow actions} coordinate a link between two TAs for capturing the flow of their behaviour;  \textbf{Terminating actions} record termination information, in addition to coordinating a link between two TAs; \textbf{Synchronisation actions} coordinate a link between a TA that participates in a synchronisation action and a TA for controlling the synchronisation;  \textbf{External choice actions} coordinate an external choice, such that choosing one of the TA that is part of the external choice thus blocking the other choices TAs; \textbf{Interrupting actions} initiate an interrupting transition that enables a TA to interrupt another; and \textbf{Exception actions} coordinate a link between a TA that raises an action for exception and a control TA that handles the action. 

\label{coordinating_action}

\end{definition}


\noindent
The names of each coordinating action are unique to ensure the correct flow of the translated TAs. In our tool, the names of the flow actions are generated in the form \lstinline{startIDx}, where \lstinline{x} is either a natural number or the name of the input \tockCSP{} process.  For instance in Figure \ref{ta_ads}, \lstinline{startID00_1} is the flow action that connects the TA00 and TA01.

Likewise, the names of the remaining coordinating actions follow the same pattern: \verb!keywordIDx!, where  \verb!keyword! is a designated word for each of the coordinating actions; \verb!finish! for a terminating action \footnote{We use terminating actions where a TA needs to communicate a successful termination for another TA to proceed. For instance, in translating sequential composition \lstinline{P1;P2}, the process \lstinline{P2} begins only after successful termination of the process \lstinline{P1}.},  \verb!ext! for an external choice action, \lstinline{intrp} for an interrupting action, and \lstinline{excp} for an exception action. Similarly, we provide a special name for a synchronising action in the form \lstinline{eventName___sync}: an event name appended with the keyword \lstinline{___sync} to differentiate a synchronising action from other actions \footnote{This is particularly important for analysis and is in the reserved keywords for the supporting tool.}.

For each translated \tockCSP{} specification, we provide an environment TA, like the TA in Figure \ref{ads_env}, which has corresponding co-actions for all the translated events of the input \tockCSP{} model, plus three coordinating actions that link the environment TA with the networked TAs. The first flow action links the environment with the first TA in the list of the translated TA (as illustrated in Figure \ref{ads_env}, the action \lstinline{startIDADS} links the environment TA with TA00 in Figure \ref{ads}). This first flow action activates the behaviour of the translated TA. Second, a terminating action to link back the terminating TA to the environment TA to capture a successful termination of a process (as shown in Figure \ref{ads_env} with the action \lstinline{FinishID0}). Third, a flow action \lstinline{tock} for recording the progress of time. A precise definition of the structure of the environment TA is as follows.




\begin{definition} 

\label{dfn_envTA}

\textbf{Environment TA} model operating environments for \uppaal. An environment TA has one state and transitions for each co-action of all the events in the input \tockCSP{} process, in addition to three transitions: the first starting flow action, the final terminating co-action and the action \lstinline{tock} for recording the progress of time.

\label{environment_TA}

\end{definition}

In translating multi-synchronisation, we adopt a centralised approach developed in \cite{O05} and implemented using Java in \cite{F05}, which uses a separate centralised controller. Here, we use a separate TA with an \uppaal{} broadcast channel to communicate synchronising information to synchronise TA and a control TA. In Figure \ref{ta_ads}, we illustrate the translation of synchronisation in translating the event \lstinline{close}, which synchronises TA03 and TA05 using the broadcasting channel \lstinline{close___sync}.





Each synchronising TA has a guard to ensure synchronisation with the correct number of TAs. The guard is a logical expression that sum of variables from all the TAs that synchronise on the synchronisation action, where each TA updates its variable from 0 to 1 to show its readiness for the synchronisation and waits for the synchronisation action. In Figure \ref{ta_ads}, the synchronising TA (TA04) has a guard expression \lstinline{(g_close00_3 + g_close01_2)==2}, which becomes true only when TA03 and TA05 update their synchronisation variables: \lstinline{g_close00_3} and \lstinline{g_close01_2}, from 0 to 1. Then, TA04 notifies the occurrence of the action \lstinline{close} and broadcasts the synchronising action \lstinline{close___sync!}. After the synchronisation, each TA resets its variable to zero and performs its remaining behaviour. A precise definition of the synchronisation TA is as follows.

\begin{definition} \textbf{A synchronisation TA} coordinates synchronisation actions. Each synchronisation TA has an initial state, and a committed state for each synchronisation action, such that each committed state is connected to the initial state with two transitions. The first transition from the initial state has a guard and an action. The guard is enabled only when all the processes are ready for synchronisation, which also enables the synchronising TA to perform the associated action that notifies the environment of its occurrence. In the second transition, the TA broadcasts the synchronisation action to all the processes that synchronise, which enables them to synchronise and proceed.
\label{dfn_syncTA}
\end{definition}






In translating external choice, we provide additional transitions to capture the behaviour of the chosen process in blocking the behaviour of the other processes. Initially, in the translated TA, all the initials \footnote{The term initials describe the first visible events of a process.} of the translated processes are available such that choosing one process blocks all the other choices.



    


\begin{figure}[t]
    \centering
    \begin{tabular}{c c}
    \includegraphics [width=0.45\textwidth]{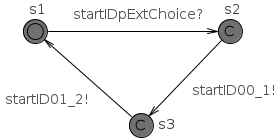} &
    \includegraphics [width=0.5\textwidth]{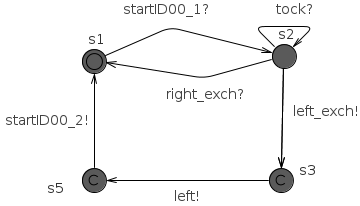} \\
    (a) TA00  & (b) TA01 \\
    \includegraphics [width=0.5\textwidth]{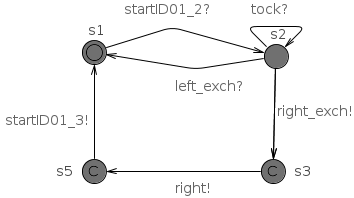} & 
     \includegraphics [width=0.3\textwidth]{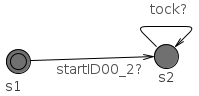} \\
    \\
    TA02 & TA03  \\
    \includegraphics [width=0.3\textwidth]{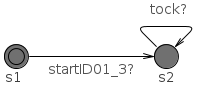} & \\
    TA04 &   \\
    \end{tabular}
    \caption{A list of TA for the translated behaviour of the process Pe}
    \label{fig_ext_choice}
\end{figure}







\begin{example} A translation of external choice is illustrated in Figure \ref{fig_ext_choice} for the process \lstinline{Pe=(left->STOP)[](right->STOP)}, which composes two processes \lstinline{left->STOP} and \lstinline{right->STOP} using the external choice operator \lstinline{([])}. 
\end{example}

\noindent
In Figure \ref{fig_ext_choice}, TA00 captures the operator external choice. TA01 and TA03 capture the LHS process (\lstinline{left->STOP}). TA02 and TA04 capture the RHS process (\lstinline{right->STOP}). TA00 has three transitions labelled with the actions: \lstinline{startIDp0_6?}, \lstinline{startID00_1!} and \lstinline{startId01_2!}. TA00 begins with the first flow action \lstinline{startIDpExtChoice?} and then starts both TA01 and TA02,  using the actions \lstinline{startID00_1!} and \lstinline{startId01_2!}, which makes them available for choice.




Initially, TA01 synchronises on \lstinline{startID00_1} and moves to location \lstinline{s2} where the TA has transitions labelled: \lstinline{left_exch?}, \lstinline{right_exch!} and \lstinline{tock?}. With the co-action \lstinline{tock?}, the TA records the progress of time and remains on the same location \lstinline{s2}. With the co-action \lstinline{right_exch?}, the TA performs an external choice co-action for blocking the LHS process when the environment chooses the RHS process, and TA01 returns to initial location \lstinline{s1}.

Alternatively, TA01 performs the action \lstinline{left_exch!} when the environment chooses the LHS process, and TA01 proceeds to location \lstinline{s3} to perform the chosen action \lstinline{left} that leads to location \lstinline{s5} and performs the flow action \lstinline{startID00_2}, which activates TA03 for the subsequent process \lstinline{STOP}. For the RHS process, TA02 captures the similar translation of the event \lstinline{right}. The omitted environment TA is similar to that in Figure \ref{ads_env}. 









In \tockCSP, a process can be interrupted by another process when composed using an operator interrupt  (\verb!/\!). Thus, we provide additional transitions to capture interruptive behaviour.







\begin{figure} [t]

    \centering

    \begin{tabular}{c c}

    \includegraphics [width=0.45\textwidth]{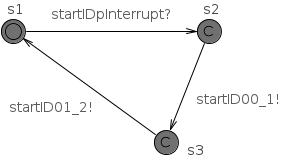} &

    \includegraphics [width=0.5\textwidth]{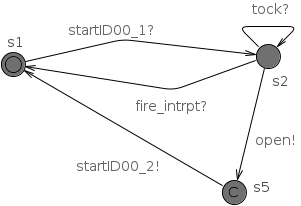}  \\

    (a) TA00 &  (b) TA01  \\

    \includegraphics [width=0.45\textwidth]{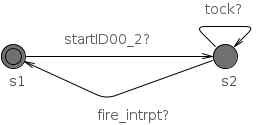} &

    \includegraphics [width=0.55\textwidth]{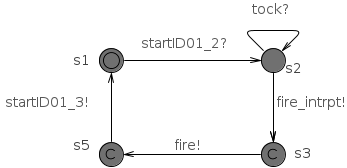} \\

    (c) TA02  &   (d) TA03  \\

    \includegraphics [width=0.45\textwidth]{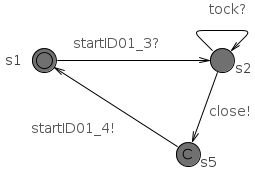}  &

    \includegraphics [width=0.4\textwidth]{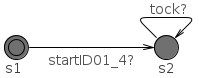} \\

    (e) TA04  & (f) TA05  \\

    \end{tabular}

    \caption{A list of TA for the translated behaviour of the process Pi.}


    \label{fig_interrupt}

\end{figure}

\begin{example} 

\label{egPI}

An example of translating interrupt is in Figure \ref{fig_interrupt}, for the translation of the process \lstinline{Pi = (open->STOP)/\(fire->close->STOP)}.
\end{example}

\noindent
In \lstinline{Pi}, the RHS process \lstinline{fire->close->STOP} can interrupt \lstinline{open->STOP} at any stable state. So, in the translated behaviour of the LHS process, we provide interruptive actions (like \lstinline{fire_intrpt}) that enable the translated behaviour of the RHS process to interrupt that of the LHS process. The corresponding co-action of the interruptive actions are provided only for the initials of the RHS process (\lstinline{fire}) because it can only interrupt with its initials. 

In Figure \ref{fig_interrupt}, TA00 is a translation of the operator interrupt, TA01 and TA02 capture the translation of the LHS process \lstinline{open->STOP}, while TA03, TA04 and TA05 capture the translation of the RHS process \lstinline{fire->close->STOP}. The environment TA is similar to the TA in Figure \ref{ads_env}.


First, TA00 performs the actions
\lstinline{startID00_1!} and \lstinline{startID01_2!} to activate TA01 and TA03. TA01 synchronises on \lstinline{startID00_1} and moves to location \lstinline{s2} where there are three possible transitions for the actions: \lstinline{tock}, \lstinline{open} and \lstinline{fire_intrpt}. With the co-action \lstinline{fire_intrpt?}, the TA is interrupted by the RHS, and returns to its initial location \lstinline{s1}. With \lstinline{tock}, the TA records the progress of time and remains on the same location \lstinline{s2}. With \lstinline{open}, the TA proceeds to location \lstinline{s5} to perform the flow action \lstinline{startID00_2} to activate TA02 for the subsequent process \lstinline{STOP}. TA02 synchronises on  \lstinline{startID00_2} and moves to location \lstinline{s2}, where it either performs \lstinline{tock?} to record the progress of time or is interrupted through the co-action \lstinline{fire_intrpt?}, and returns to its initial location \lstinline{s1}.



For the RHS, TA03 captures the translation of the event \lstinline{fire}. TA03 begins with synchronising on \lstinline{startID01_2}, which progresses by interrupting the LHS process using the interruptive flow action \lstinline{fire_intrpt}, then \lstinline{fire}, and performs \lstinline{startID01_3} for activating TA04 which synchronises on the flow action and moves to location \lstinline{s2}, where it either performs the action \lstinline{tock?} for the progress of time and remains in the same location or performs the action \lstinline{close} and proceeds to location \lstinline{s5}, then performs the flow action \lstinline{startID01_4} for starting TA05 for the translation of \lstinline{STOP} (deadlock).





We translate  \lstinline{tock} into a corresponding action \lstinline{tock} using a broadcast channel for the environment TA to broadcast the progress of time for all the TAs to synchronise. For instance, in Figure \ref{ta_ads}, the environment TA has a transition labelled \lstinline{tock} guarded with the clock expression $ck \geq 1$, so that \lstinline{tock} happens every 1 time unit, and resets the clock $ck=0$ to zero on following the transition.

Also, we translate non-deterministic choice into silent transitions, such that the translated TA follows one of the silent transitions non-deterministically. This completes an overview of the strategy we follow in developing the translation technique. A precise description of all the translation rules in Haskell is in \cite{extendedVersion, thesis}.

\section{Evaluation}

\label{evaluation}


A sound translation ensures that the properties of the source model are preserved in the translated model. This is determined by comparing their behaviours \cite{T06, NN07, KM19, B81}. We compare the behaviour of the input \tockCSP{} and the output TA in two phases: experimental evaluation and mathematical proof.

\subsection{Experimental Evaluation}

\begin{figure}[t!] 
    \centering
    \includegraphics[width=\textwidth, height=0.25\textheight]{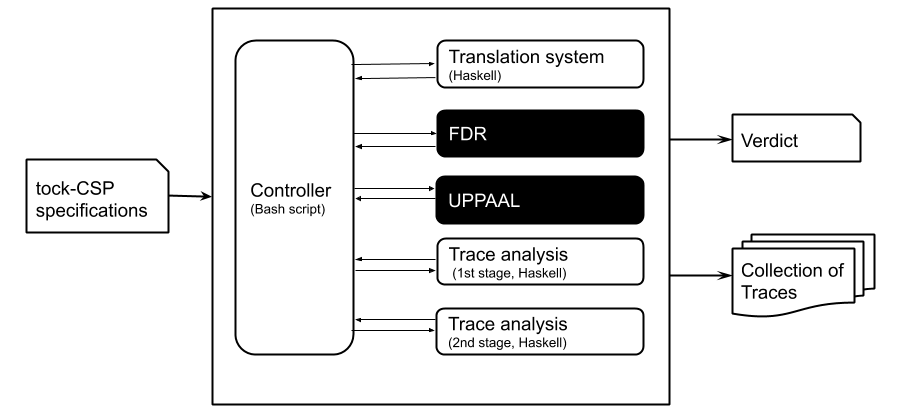}
    \caption{Structure of the trace analysis system}
    \label{trace_analysis_system}
\end{figure}

In comparing the traces, we use trace semantics, we have developed an evaluation tool, which uses our translation tool and both FDR and \uppaal{} as black boxes for generating finite traces. The structure of this tool \footnote{Available for download  at \url{https://github.com/ahagmj/TemporalReasoning.git}} is in Figure \ref{trace_analysis_system}.



In generating traces, like most model checkers, FDR produces only one trace (counterexample) at a time. So, based on the testing technique in \cite{NS08}, we have developed a trace-generation technique that repeatedly invokes FDR until we get all the required trace sets of the input process. Similarly, based on another testing technique with temporal logic \cite{LP07}, we have developed a trace-generation technique that uses \uppaal{} to generate traces of the translated TA models. 


These two trace-generation techniques form components of our trace analysis tool (Figure \ref{trace_analysis_system}), which has two stages. In the first stage, we generate traces of the input \tockCSP{} and its corresponding translated TA, using both FDR and \uppaal. Then, we compare the generated traces; if they do not match, we move to the second stage where we use FDR to complement \uppaal{} in generating traces. The mismatch may be because FDR distinguishes different permutations of events (traces). In contrast,  \uppaal{}  uses a logical formula to generate traces \cite{LP07, BD06} which do not distinguishes traces with different permutations.




Essentially, \uppaal{} checks if a system satisfies its requirement specifications (logical formula), irrespective of the behaviour of the system. For example, \uppaal \ does not distinguish between the two traces $ \langle e1, e2, e3 \rangle$ and  $\langle e1, e3, e2 \rangle$, if both traces satisfy the requirement specification formula. However, FDR is capable of generating both traces. Thus, in the second stage, we use \uppaal{} to check if all the traces of FDR are acceptable traces of the translated TA.




For evaluation, we have used a list of systematically formulated \tockCSP{} processes that pair the constructs of \tockCSP{}. The list contains 111 processes. Archives of the processes and their traces are available in this repository \cite{extendedVersion}.





In addition, we test the translation technique with larger examples from the literature, such as an automated barrier to a car park \cite{S10}, a thermostat machine for monitoring ambient temperature \cite{S10}, an Automated Teller Machine (ATM)  \cite{R05}, a bookshop payment system \cite{S10}, and a railway crossing system \cite{R10}. An overview of these case studies is in Table \ref{table_caseStudies}, while the details, including the traces, are also available in the repository of this work \cite{extendedVersion}.

Considering that the experimental approach with trace analysis is an approximation for establishing correctness with a finite set of traces, covering infinite sets of traces in proving correctness has to use mathematical proof. 





{ \renewcommand{\arraystretch}{1.2}
\setlength{\tabcolsep}{10pt}
\begin{table}[t!]
    \centering
    \caption{An overview of the case studies}
    \label{table_caseStudies}
    \begin{tabular}{| c | c | c | c | c |}  \hline \hline
No. & System &    States  &    Transitions     &   Events     \\ \hline \hline

1 & Thermostat Machine      &      7    &   16            &   5       \\   \hline  \hline

2 & Bookshop Payment System &     7    &   32            &    9       \\   \hline \hline
3 & Simple ATM             &     15  &    33           &  15         \\ \hline \hline
4 & AutoBarrier system     &     35    &   84             &  10        \\  \hline \hline
5 & Rail Crossing System &     80    &   361            &   12       \\   \hline \hline

    \end{tabular} 

\end{table}

}


%

\subsection{Mathematical Proof}







A more detailed account of our proof can be found in \cite{extendedVersion, thesis}. Here, we illustrate part of the proof using one of the base cases of the structural induction. Mathematically TA is defined as tuple \footnote{TA = $(L, l_0, C, A, E, I)$ where $L$ is a set of locations, $l_0$ is the initial location, $C$ is a set of clocks, $A$ is a set of actions, $E$ is a set of edges and $I$ is an invariant.}  (Section \ref{background}). Consider \lstinline{TA1} as the translation of the process \lstinline{STOP} (\lstinline{TA05} from Figure \ref{fig_interrupt}), then mathematically TA1 is expressed as follows:











\vspace{-0.7cm}
\begin{multline}
TA1 =   ( \{s1, s2\}, s1,  \{ck\}, \{startID01\_4, tock\}, \\ 
        \{(s1, startID01\_4, \emptyset, \emptyset, s2), (s2, tock, ck\le1, ck, s2) \}, \{\} )
\end{multline}













\noindent
In the language of TA, a path \cite{AD94, B11} is a sequence of consecutive transitions that begins from the initial state. A trace \cite{AD94, B11} (or word) is a sequence of actions in each path. In \lstinline{TA1}, there is only one infinite path; the first transition from location $s1$ to location $s2$ and the second transition from location  $s2$ and return to location $s2$, repeated infinitely ($s2$ to $s2$). The traces on the path are as follows:


\vspace{-0.5cm}
\begin{equation}
traces`_{TA}(TA1) =  \{  \langle \rangle \}  \cup   \{ \langle startID01\_4 \rangle ^ \frown  \langle tock  \rangle ^ n \mid n \in N  \} 
\end{equation}
\vspace{-0.2cm}




\noindent
The function $trace`_{TA}(TA)$ describes the traces of TA1 as follows: the first empty sequence happens at the initial state, before the first transition; the action \lstinline{startID01_4} happens on the first transition; the action \lstinline{tock} happens on the second transition, which is repeated infinitely for the infinite traces $\langle tock  \rangle ^ n $.










Another function $trace_{TA}(TA)$ is similar to $traces`_{TA}(TA)$ but removes all the coordinating actions (Definition \ref{coordinating_action}) from the traces.






\vspace{-0.5cm}
\begin{equation}
traces_{TA}(TA) = \{ t \ \backslash  \ CoordinatingActions \mid t \in traces`_{TA}(TA)  \}
\end{equation}




\noindent
Therefore, without coordinating actions, the traces of TA1 become.
 \vspace{-0.1cm}
\begin{equation}
traces_{TA}(TA1) = \{ \langle tock \rangle ^ n  \mid n \in N  \}
\end{equation}














\vspace{-0.2cm}

\begin{proof} For this proof, our goal is to establish that the traces of \tockCSP{} models are the same as those of the translated TA models. Here, \lstinline{transTA} is the translation function we have formalised (see Section \ref{trans_rule}) for translating \tockCSP{} models into TA models.  Thus, for each valid \tockCSP{} process \lstinline{P}, within the scope of this work, we need to establish that:
\end{proof}

 \vspace{-0.5cm}
\begin{equation}
traces_{tock-CSP}(P) = traces_{TA}(transTA(P)) 
\end{equation}




\noindent
Therefore, for each translation rule, we have to prove that the translated TAs capture the behaviour of the corresponding input \tockCSP{} model. We use automatic proof, expressed in Haskell. Starting with the basic process \lstinline{STOP}:







\begin{lstlisting}[numbers=left,numbersep=7pt, language=Haskell, breaklines=true, basicstyle=\ttfamily\small ]
         traces_tockCSP(STOP) = traces_TA(transTA STOP)
\end{lstlisting}


\noindent
Using structural induction in Haskell, we show that:
\begin{lstlisting}[numbers=left,numbersep=7pt, language=Haskell, breaklines=true, basicstyle=\ttfamily\small ]
(traces_tockCSP n STOP = traces_TA n (transTA STOP)) 
 => (traces_tockCSP (n+1) STOP = traces_TA (n+1) (transTA STOP))
\end{lstlisting}

\noindent
Each step is evaluated automatically that helps us prove that the traces of the translated TAs capture the traces of the input \tockCSP{} correctly. Detailed steps of the proof are available in the extended reports \cite{thesis, extendedVersion}.

\section{Related Work}
\label{related_work}


Timed-CSP \cite{S10} is another popular extension of CSP for capturing temporal specifications. Unlike \tockCSP, Timed-CSP records the progress of time with a series of positive real numbers. However, the approach of Timed-CSP can not specify deadline nor urgency. Also, traces of Timed-CSP are infinite, which is problematic for automatic analysis and verification \cite{R10}. Thus, there is no tool support for verifying Timed-CSP models. As a result of this, many researchers explore model transformations in translating Timed-CSP into \tockCSP{} for using FDR in automatic verification \cite{O00}. Also, Timed-CSP has been translated into \uppaal, initially reported in \cite{DH08} and then subsequently improved in \cite{TS13}. Additionally, Timed-CSP has been translated into Constraint Logic Programming (CLP)  for reasoning with the constraint solver CLP(R) \cite{DP06}.



However, there is less focus on applying the same transformation techniques for \tockCSP{}. Although, an attempt to transform TA into \tockCSP{} was proposed in \cite{K11}, whereas in this work, we consider the opposite direction.




Apart from CSP and TA, model transformations have been used for improving various formal modelling notations. For instance, Circus has been translated into \lstinline{CSP||B} for using the tool ProB for automatic verification \cite{KJ17}. Additionally, the language B has been translated into TLA+ for automatic validation with TLC \cite{HL14}. Also, translating TLA+ to B has been investigated for automated validation of TLA+ with ProB \cite{HL12}, such that both B and TLA+ benefit from the resources of each other, and their supporting tools ProB and TLC, respectively.


Model transformation has become an established field for addressing computational problems; here we also consider model transformation. The novelty of this translation work is traced back to the early days of model translation works, such as the translation of the timed variants of LOTOS into TA \cite{CO95, HG98}. A recent systematic survey of model transformation provides a rich collection of model transformation techniques \cite{KM19}.




\section{Conclusion}
\label{conclusion}


In this work, we have presented a technique for translating \tockCSP{} into TA for \uppaal{} to facilitate using temporal logic and facilities of \uppaal{} in verifying \tockCSP{} models. This work contributes an alternative way of using TCTL to specify liveness requirements and other related requirements that are difficult to verify in \tockCSP{} with refinement. Also, our work sheds light into the complex relationship between \tockCSP{} and TA (temporal logic model).


Currently, we translate the event \lstinline{tock} into an action that is controlled by a timed clock in \uppaal.  A recommended next step is to relate the notion of \lstinline{tock} to the notion of time in TA and get rid of \lstinline{tock} as an action.  This additional extension will help us to explore additional interesting facilities of \uppaal{} to verify temporal specifications. Also, in future work, a better understanding of relating \tockCSP{} to TA will help us to explore using a single TA instead of network TAs for more efficient verification.



\bibliography{ref}
\bibliographystyle{splncs04}

\end{document}